\documentclass{moriond}

\def\Journal#1#2#3#4{{#1} {\bf #2}, #3 (#4)}

\def\PRL{\em Phys. Rev. Lett.}
\def\PRD{{\em Phys. Rev.} D}

\def\IBD{\bar \nu_e + p \rightarrow e^+ + n}

\def\be{\begin{equation}}
\def\ee{\end{equation}}
\def\bea{\begin{eqnarray}}
\def\eea{\end{eqnarray}}

\usepackage{amsmath}
\usepackage{amssymb}
\usepackage{xspace}
\newcommand{\ket}[1]{\ensuremath{|#1\rangle}\xspace}

\usepackage[squaren,Gray]{SIunits}

\newcommand{\Photo}{\includegraphics[height=35mm]{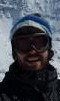}}

\begin{document}
\vspace*{4cm}
\title{Search for a sterile neutrino with the STEREO detector at ILL}

\author{ S. ZSOLDOS }

\address{Laboratoire de Physique Subatomique et Cosmologie (LPSC), 53 Avenue des Martyrs,\\
38026 Grenoble Cedex, France}

\maketitle\abstracts{
In 2011, a re-evaluation of the antineutrino spectrum emitted by nuclear reactors revealed a 6\% deficit between the observed flux and the expected one. This anomaly is significant at 2.9$\sigma$ and can be explained by a new oscillation at short range due to a light sterile neutrino, with parameter $\Delta m^2$ = 0.1-1 eV. The STEREO detector in construction at ILL will be the first ever to measure with precision the antineutrino spectrum and flux at very short distance (9-11m) from a compact reactor core and it will be able to confirm or reject the existence of this light sterile neutrino. In the following we introduce the relevant parameter to study the neutrino oscillation, then the STEREO detector and its discovery potential.
}

\section{Status of neutrino mixing and anomalies}
	\subsection{Introduction}
Since the sixties and shortly after the neutrino was discovered, several experiments reported anomalous values of the flux from different sources of neutrinos (so-called \textit{solar}, \textit{atmospheric} and \textit{reactor}), which were smaller than expected.

Convincing evidence was finally smeared out in 1998 by the Super-Kamiokande Collaboration~\cite{SuperK1998}, which was looking at neutrinos created in the atmosphere.
From then on, many other experiments validated this observation and accurately probed the solar (SNO, GALLEX, SAGE, Super-K) and the reactor (KamLAND, CHOOZ, Palo Verde, Daya Bay, Reno, Double CHOOZ) sectors as well.
Moreover, other experiments (LSND, miniBOONE) measured the apparition of electronic neutrino in a pure muonic neutrino beam produced by an accelerator source.

These results are interpreted within the framework of oscillation and comfort the hypothesis of a non-vanishing mass for neutrinos.
To produce oscillation in the Standard Model, we assume that mass eigenstates, which are the ones who propagates through Shrodinger equation, differ from interaction eigenstates.
We can therefore define a change-of-basis matrix $U$ by :

	\begin{equation}
	\label{uPMNS}
	\underset{\alpha=e,\mu,\tau}{\ket{\nu_\alpha}} = \sum_{i=1}^3 U_{\alpha i}\ket{\nu_i} \Longleftrightarrow
		\underset{\text{flavor}}{
			\begin{pmatrix}
  			\nu_{e} \\
  			\nu_{\mu} \\
  			\nu_{\tau} \\
			\end{pmatrix}
		}
			=
 			\begin{pmatrix}
  			U_{e1}  & U_{e2}  & U_{e3} \\
  			U_{\mu 1} & U_{\mu 2} & U_{\mu 3} \\
  			U_{\tau 1} & U_{\tau 2} & U_{\tau 3} \\
			\end{pmatrix}
		\underset{\text{mass}}{
			\begin{pmatrix}
  			\nu_{1} \\
  			\nu_{2} \\
  			\nu_{3} \\
			\end{pmatrix}
		}
	\end{equation}

The transition probability between two flavors states is given by computing the squared amplitude of the wave function :

	\begin{equation}
	\label{prob}
	P(\alpha\rightarrow\beta) = \Vert A(\alpha\rightarrow\beta) \Vert^2 = \sum_i \sum_j U_{\alpha i} U_{\beta i}^{-1} \overline{U_{\alpha j} U_{\beta j}^{-1}} \exp{\left(-i\Delta m_{ij}^2\frac{L}{2E}\right)}
	\end{equation}

From Eq. \ref{prob}, a non-zero mass for neutrinos has to be assumed in order to observe the oscillation mechanism.
The measured values of the squared neutrino mass eigenstate difference are $\Delta m_{12}^2$ = 7.54 $\pm$ 0.26 $\times$ 10$^{-5}$ eV$^2$ and $\Delta m_{13}^2$ $\simeq$ $\Delta m_{23}^2$ = 2.43 $\pm$ 0.06 $\times$ 10$^{-3}$ eV$^2$.

The matrix $U$ is unitary, and so can be parametrized by three euler angles.
This framework corresponds to the Pontecorvo-Maki-Nakagawa-Sakata matrix, and is given by :

	\begin{equation}
			U_\text{PMNS} =
			\begin{pmatrix}
  			1  & 0  & 0 \\
  			0  & \cos \theta_{23}  & \sin \theta_{23} \\
  			0  & -\sin \theta_{23}  & \cos \theta_{23}
			\end{pmatrix}
			\times
			\begin{pmatrix}
  			\cos \theta_{13}  & 0  & \sin \theta_{13} \\
  			0  & 1  & 0 \\
  			-\sin \theta_{13}  & 0  & \cos \theta_{13}
			\end{pmatrix}
			\times
			\begin{pmatrix}
  			\cos \theta_{12}  & \sin \theta_{12}  & 0 \\
  			-\sin \theta_{12}  & \cos \theta_{12}  & 0 \\
  			0  & 0  & 1
			\end{pmatrix}
 	\end{equation}

Historically, we define the mixing angle with respect to the region were the oscillation is driven, and the corresponding angle between mass states are $\theta_{23}$ = $41.4\degree^{+1.9}_{-1.3}$ $:$ \textit{athmospheric} (1998), $\theta_{12}$ = $33.7\degree^{+1.1}_{-1.1}$ $:$ \textit{solar} (2001), $\theta_{13}$ = $8.80\degree^{+0.37}_{-0.38}$ $:$ \textit{reactor} (2012).
So far, this three neutrinos mixing scheme gives a very good fit to most of measurements~\cite{Lisi2015} except for some anomalies among which the reactor antineutrino anomaly that is discussed below is found.

	\subsection{Recent results}

\begin{figure}[htb]
\begin{center}
	\includegraphics[scale=0.07]{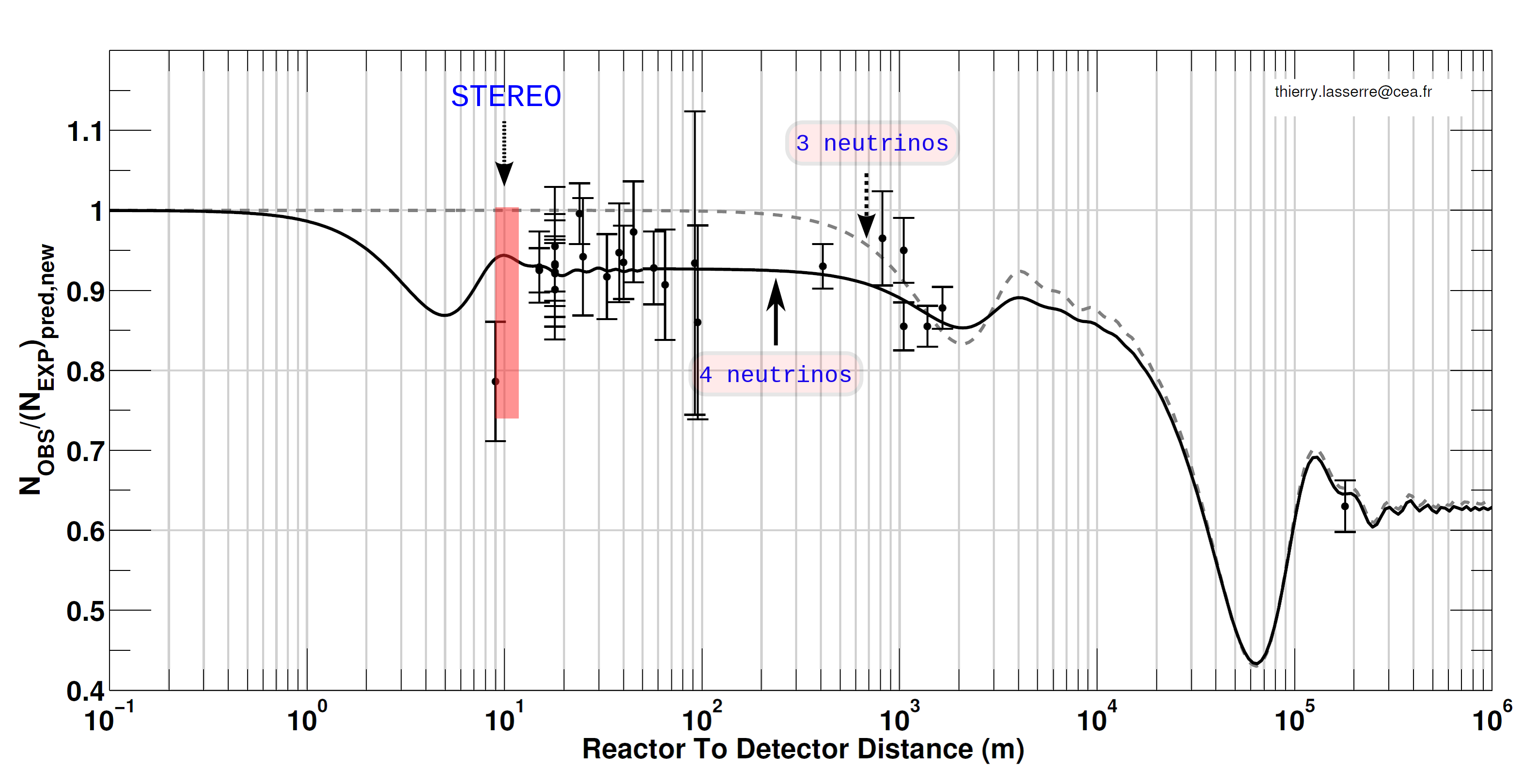}
	\includegraphics[scale=0.27]{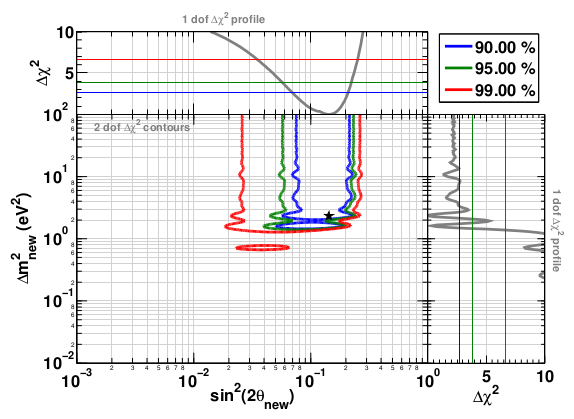}
\caption{Left : Reevaluation of the reactor antineutrinos experiment with the 3 neutrinos fit (dashed) and with an additional sterile state (solid). Right : Allowed regions of $\sin^2(2\theta_{new})$ and $\Delta m_{new}^2$ by the reactor antineutrino anomaly}
	\label{RAA}
\end{center}
\end{figure}

In 2011, a reevaluation of the electron antineutrino flux	 from nuclear reactor suggested an increase of the expected flux with respect to previous estimations~\cite{Mueller2011}.
This study led to a reanalysis of previous experiments~\cite{Mention2011} and showed a deficit of 2.9$\sigma$ from the expected flux.
When combined with previous \textit{Gallium anomaly}~\cite{Giunti2010} (where a deficit of neutrinos was observed during the calibration of \textsf{Ga} detectors - typically for solar experiments), the effect becomes significant to 3.6$\sigma$.
This deviation is shown on Fig. \ref{RAA}.

One possible explanation for this observation could be the existence of a new neutrino state, which oscillates with the other ones.
This state has to be \textit{sterile}, because of the Z decay width measurement at LEP, which constraints the number of states interacting with the Z boson to $2.9840\pm0.0082$~\cite{LEP2005}.
The best fit parameters for this new oscillation are $\sin^2(2\theta_{new})=0.17\pm0.04$(1$\sigma$) and $\Delta m_{new}^2=2.3\pm0.1$(1$\sigma$).

Due to the high value of $\Delta m_{new}^2$, of the order of one eV, one of the best way to search for this sterile state is by looking at the disappearance of reactor antineutrinos at very short distance from their source, where the probability of oscillation is maximum.
Note that all of the previous very short baseline experiments measured the antineutrino flux at some distance and compared it to the expected calculated flux.
These "flux only" measurements are subjected to uncertainties on the energy spectrum of the emitted antineutrinos by reactors~\cite{Hayes2015}.
To work around this problem and to sign unequivocally the oscillating character of the propagation, we are proposing instead to observe the deformation of the interacting neutrino energy spectrum as a function of the distance from the source.


\section{The STEREO detector}

The STEREO experiment aims at measuring the rate and energy spectrum of antineutrinos emitted by the ILL reactor in Grenoble, France, at 6 different positions between 9m and 11m from the core.
Antineutrinos are detected through their inverse beta decay reaction (IBD) : $\IBD$.
The experimental signature is based on a delayed coincidence between the positron annihilation and the neutron capture.


The detector consist of a 6 cells target volume of $40\times40\times90$ cm$^3$, each filled with a liquid scintillator doped with gadolinium, to enhance the neutron capture.
The positron deposits almost instantanously its energy, which constitutes the \textit{prompt} signal.
This prompt visible energy in the detector is defined as the kinetic energy from the positron plus its anihilation with an electron :
\begin{equation}
	E_\text{vis} = E_{\bar\nu_e} + m_p - m_n + m_e = E_{\bar\nu_e} - 0.782 \text{MeV}
\end{equation}
The neutron created by the IBD process with a few 10 keV energy will rapidly thermalize in the liquid scintillator and then it will diffuse for few $\mu$s until being captured on a nucleus.
Doping the liquid with gadolinium allows to reduce the diffusion time and to sign the neutron capture by a 8 MeV $\gamma$-cascade well above the natural background.

\begin{figure}[htb]
\begin{center}
	\includegraphics[scale=0.32]{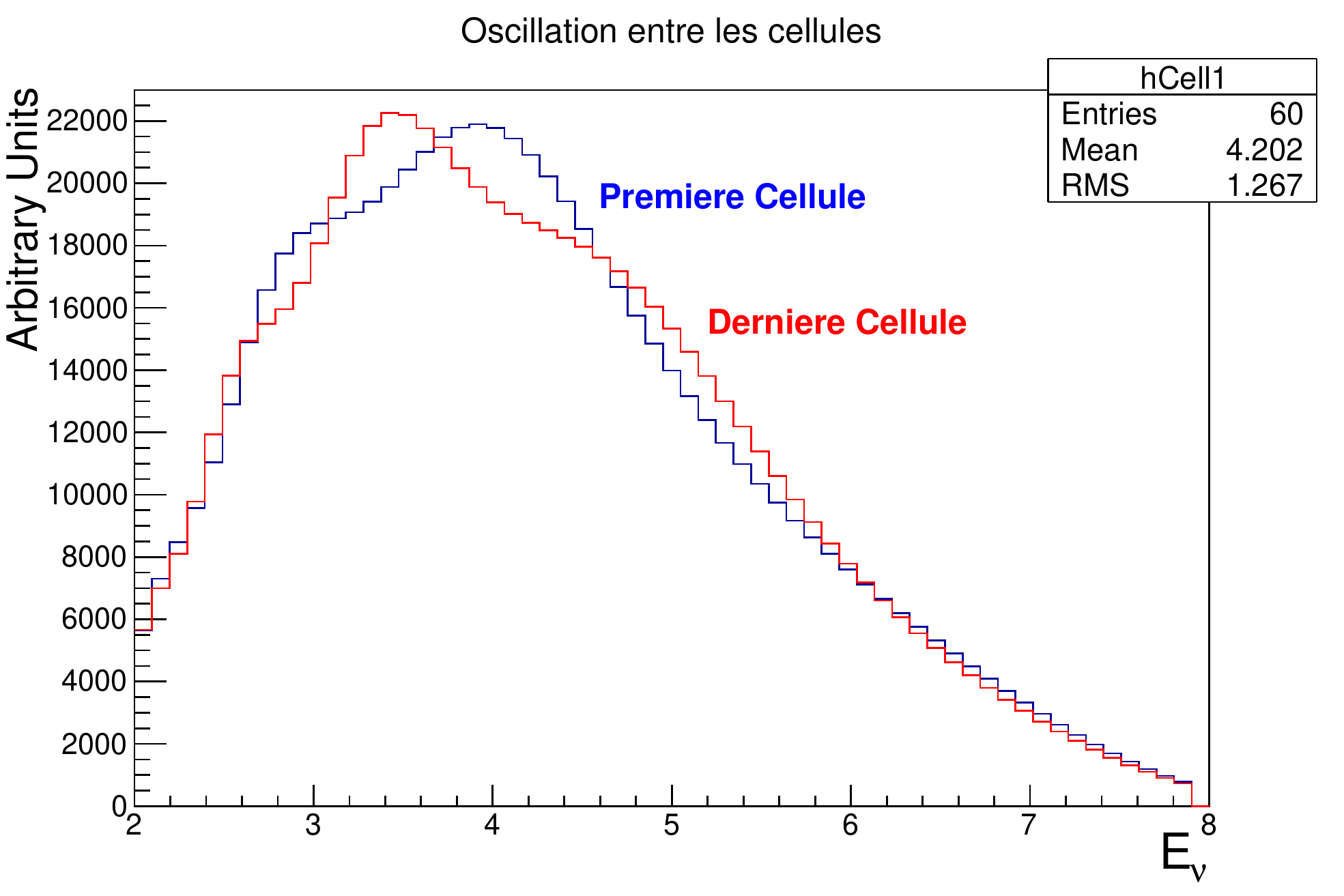}
	\includegraphics[scale=0.30]{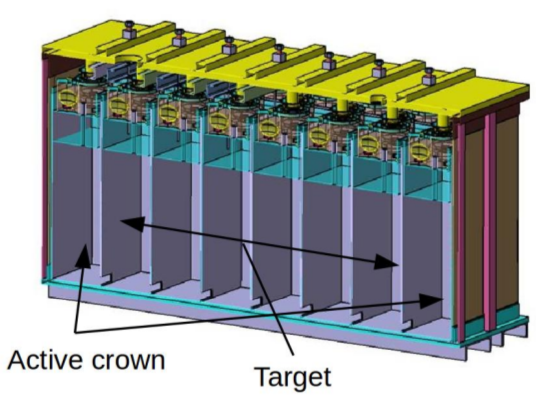}
	\caption{Left : Antineutrinos energy spectrum deformation between the first cell (in blue) and the last one (in red), in case of a new oscillation at the best fit parameters. Right : Scheme of the STEREO detector, where the 6 targets and $\gamma$-catcher are represented. Photomultipliers are placed above the cells and separated by a acrylic buffer to enhance the signal. The top is covered by the water-cherenkov muon veto}
	\label{designStereo}
\end{center}
\end{figure}

The target volume is placed between two cells filled with liquid scintillator only, called the $\gamma$-catcher.
It aims at collecting the energy deposited by events happening at the edge of the target : they might create gamma rays which will escape the target volume, and this would bias the energy reconstruction of the event if missed.

On the top of the cells, a thick acrylic buffer separates the liquid scintillator from the photomultipliers.
Its aim is twofold : optimizing the detection efficiency and increasing background rejection and shielding.

Covering the whole surface of the target volume and $\gamma$-catcher, a water-cherenkov detector aimed at detecting cosmic muons will work as a veto for the acquisition.

\section{Sensitivity and discovery potential}

The challenge raised by measuring the energy spectrum of antineutrinos close to a reactor core is mostly to control multiple sources of background.
They are divided into two categories : the accidental background, where two random processes emit light in the detector in a time window of a few $\mu$s, and the correlated background, where only one process mimick the IBD signal.

\subsection{Accidental background}
The main sources of accidental background come from thermal neutrons and $\gamma$ produced in the surroundings of the detector.
Campaigns of measurements have been performed to characterize the sources and the energies of those backgrounds.
Heavy shieldings of lead (against $\gamma$) and polytehylene (against neutrons) have been designed and are being placed all around the detector.
The expected rate of accidental events is defined as :
\begin{equation}
	R_\text{accidental} = \Phi_\text{n} \times \Phi_\gamma \times T_\text{window}
\end{equation}
Thanks to our heavy shielding, we expect to drop thermal neutrons and $\gamma$ rate to $\Phi_\text{n} \simeq 10\text{Hz}$ and $\Phi_\gamma\simeq 10\text{Hz}$.
With $T_\text{window}\simeq 10\mu\text{s}$, we expect less than 1mHz of accidental backgrounds.

\subsection{Correlated background}
Essentially all of the correlated background comes from fast neutrons interacting in the detector.
They hit protons whose recoil induces a scintillation light (mimick the prompt signal), and they are captured shortly afterwards (mimick the delayed signal).
Sources of fast neutrons are the reactor core and cosmic muons.
A high-energy muon can interact with a high-Z material and create fast neutrons by spallation.
Additional polyethylene shielding is used to moderate fast neutrons from the reactor and the veto above the detector allows to tag muons events.
Thanks to these additional shielding, the expected rate of fast neutrons inside the detector is expected to be at the order of $\simeq$ 1mHz.

\subsection{Sensitivity}

\begin{figure}[htb]
\begin{center}
	\includegraphics[scale=1.]{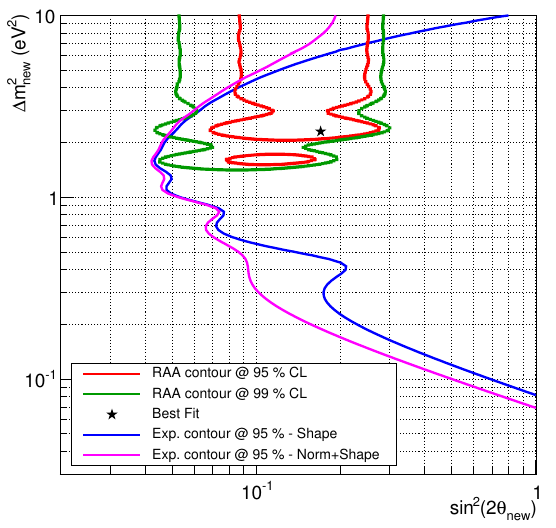}
	\includegraphics[scale=0.26]{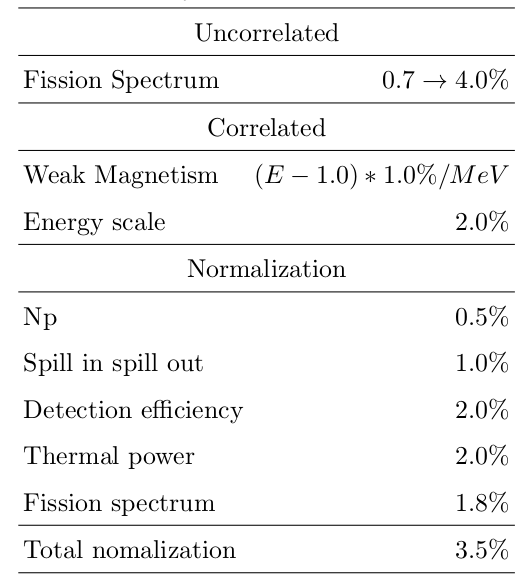}
	\caption{Sensitivity of the STEREO experiment. Area at the right of the blue and pink contours are the region explored by the detector after 300 days of data taking at 95\% CL, respectively with the energy shape distorsion analysis only and with the shape + norm analysis. The red and green contours are the region allowed by the reactor antineutrino anomaly at 95\% CL and 99\% CL respectively, and the star shows the best fit parameters. On the right, the table summing the uncertainties effect taken in account}
	\label{sensitivity}
\end{center}
\end{figure}

The expected sensitivity of the experiment is shown in Fig. \ref{sensitivity}.
The STEREO detector will be able to cover all the reactor antineutrino anomaly in 2 years of data taking with a 95\% CL.
Detection and reconstruction of systematics are included, as well as systematics of the antineutrino spectrum.

\end{document}